# Seeing moiré superlattices

L. J. McGilly[1], A. Kerelsky[1], N. R. Finney[2], K. Shapovalov[3], E.-M. Shih[1], A. Ghiotto[1], Y. Zeng[1], S. L. Moore[1], W. Wu[4], Y. Bai[4], K. Watanabe[5], T. Taniguchi[5], M. Stengel[3,6], L. Zhou[4,7], J. Hone[2], X.-Y. Zhu[4], D. N. Basov[1], C. Dean[1], C. E. Dreyer[8,9], and A. N. Pasupathy[1,*]

[1]Department of Physics, Columbia University, New York, New York 10027, USA
[2]Department of Mechanical Engineering, Columbia University, New York, NY 10027, USA
[3]Institute of Materials Science of Barcelona, Bellaterra 08193, Spain
[4]Department of Chemistry, Columbia University, New York, NY 10027, USA
[5]National Institute for Materials Science, Tsukuba, Japan
[6]ICREA–Instució Catalana de Recerca i Estudis Avançats, 08010 Barcelona, Spain
[7]College of Engineering and Applied Sciences, Nanjing University, Nanjing 210093, China
[8]Department of Physics and Astronomy, Stony Brook University, Stony Brook, NY 11794, USA
[9]Center for Computational Quantum Physics, Flatiron Institute, 162 5th Avenue, New York, NY 10010, USA
*e-mail: apn2108@columbia.edu

**Moiré superlattices in van der Waals (vdW) heterostructures have given rise to a number of emergent electronic phenomena due to the interplay between atomic structure and electron correlations. A lack of a simple way to characterize the local structure of moiré superlattices has impeded progress in the field. In this work we outline a simple, room-temperature, ambient method to visualize real-space moiré superlattices with sub-5 nm spatial resolution in a variety of twisted vdW heterostructures including but not limited to conducting graphene, insulating boron nitride and semiconducting transition metal dichalcogenides. Our method utilizes piezoresponse force microscopy, an atomic force microscope modality which locally measures electromechanical surface deformation. We find that all moiré superlattices, regardless of whether the constituent layers have inversion symmetry, exhibit a mechanical response to out-of-plane electric fields. This response is closely tied to flexoelectricity wherein electric polarization and electromechanical response is induced through strain gradients present within moiré superlattices. Moiré superlattices of 2D materials thus represent an interlinked network of polarized domain walls in a non-polar background matrix.**

A distinguishing feature of 2D materials is that they can be easily stacked atop each other regardless of lattice parameters and orientation in stark contrast to other condensed matter heterostructures where epitaxial coherence is required to produce pristine interfaces. However, this greater flexibility is accompanied by added complexity, as the moiré superlattice resulting from the lattice mismatch between the stacked layers gives rise to a spatially varying periodic structural and potential landscape that can dramatically alter the electronic bandstructure. Indeed, electrons in these structures have been recently found to exhibit a number of emergent properties that the individual layers themselves do not exhibit. This includes superconductivity [1,2], magnetism [3], topological edge states [4,5], exciton trapping [6] and correlated insulator phases [7]. However, even though the electronic bandstructure can be exquisitely sensitive to the geometry of the moiré pattern, characterizing this geometry independently remains a difficult tast. In fact, the same flexibility that allows for the creation of heterostructures without epitaxy implies that these structures are highly susceptible to heterostrain and interfacial impurities that modify or disrupt the moiré periodicity. Existing methods to visualize the moiré superlattices with high resolution, including transmission electron microscopy [8,9] and scanning tunneling microscopy [10] require some combination of ultra-high vacuum, low temperature, complex setups or specialized sample preparation that makes these methods impractical to apply on a regular basis to characterize moiré structures. Other methods such as near-field optics [5] and transport with multiple contacts [1-3,7] are limited in resolution to length scales above the moiré periods of interest, which are typically in the 10 nm size scale. There is thus an urgent need for a facile method to characterize moiré superlattices in samples of interest.

In this work we show that moiré superlattices formed between two arbitrary 2D layers have an electromechanical response to applied electric fields. We leverage this universal effect to directly image moiré structures with sub-5 nm resolution in a commercial atomic force microscope (AFM), operated in the piezoresponse force microscopy (PFM) mode. This simple method works in ambient atmosphere at room-temperature without the need for complex sample preparation or elaborate device fabrication. The surprisingly universal electromechanical response of moiré structures is explained through the flexoelectric effect, where strain gradients present in moiré superlattices produce an electric polarization. Our results show that moiré superlattices are universally composed of a polar array of 1D channels and nodes in a non-polar background matrix.

We first focus on graphene, the best understood of the 2D materials, as a prototypical model system before expanding to other vdW heterostructures. One of the simplest multilayer vdW systems is twisted bilayer graphene (tBLG); a stack of two monolayers of graphene with different lattice orientations, as can be seen in Figure 1a-c. Despite this simplicity, there are still a number of ways in which the two layers can be arranged. These stacking configurations are denoted by the vertical layout of the two sublattices such that "AA" stacking identifies regions where the A (B) sublattice in one layer is directly above the corresponding A (B) sublattice of the lower layer (Figure 1a). This is a highly energetically unfavourable configuration compared to "Bernal" or "AB" stacking (Figure 1b), so there is considerable energetic drive for the system to uniformly adopt the AB (or BA) stacking arrangement. The configuration at the interface between regions of AB and BA stacking is called "saddle point" (SP) stacking, as seen in Figure 1c, and again incurs an energetic cost above pure AB stacking. The rotational misalignment of layers enforces the creation of a moiré superlattice that will necessarily contain each of the stacking domains in proportions that are related to the twist angle. Below a critical angle [11] there is an energetic drive to form a moiré pattern that maximizes the energetically favourable AB stacking at the expense of AA sites through lattice reconstruction which leads to the creation of discrete stacking domains and domain walls [8,9]. For bilayers with zero twist angle the moiré pattern will instead result from a lattice mismatch due to either dissimilar materials or a differential strain (heterostrain) between the two layers.

As a first step, we begin with low-twist angle tBLG before generalizing our results to a wide range of materials. Figure 1e shows a schematic of the PFM technique, in which an AC bias applied between the tip and sample induces periodic deformation of the sample whose amplitude and phase provide information about the local electromechanical response (see Methods). As shown in Figure 1g,h, PFM produces amplitude and phase images which clearly show the domain wall array of the tBLG moiré pattern. Simple inspection of the amplitude shows that a large response is found at the domain wall regions and at the AA stacking sites, while a smaller but nonzero response is found in the AB domains. This is an interesting but somewhat surprising result as from a simple symmetry perspective there is no reason to believe that bilayer graphene, which is centrosymmetric, would be electromechanically active through the piezoelectric effect that is usually probed by PFM (there are a number of studies that suggest graphene can be piezoelectric given certain conditions but these are not applicable here [12-17]). It should be noted that the height image (Figure S1) does not show any apparent topography, within the noise levels of the instrument ruling out crosstalk from large surface features. However, PFM can give responses that on first glance seem to indicate piezoelectric or even ferroelectric behavior but which on closer inspection turn out to be electrostatic [18] and electrochemical effects [19] or even cantilever dynamics [20]. Therefore a careful analysis is required to understand the nature of the contrast obtained in the images of Figure 1g,h.

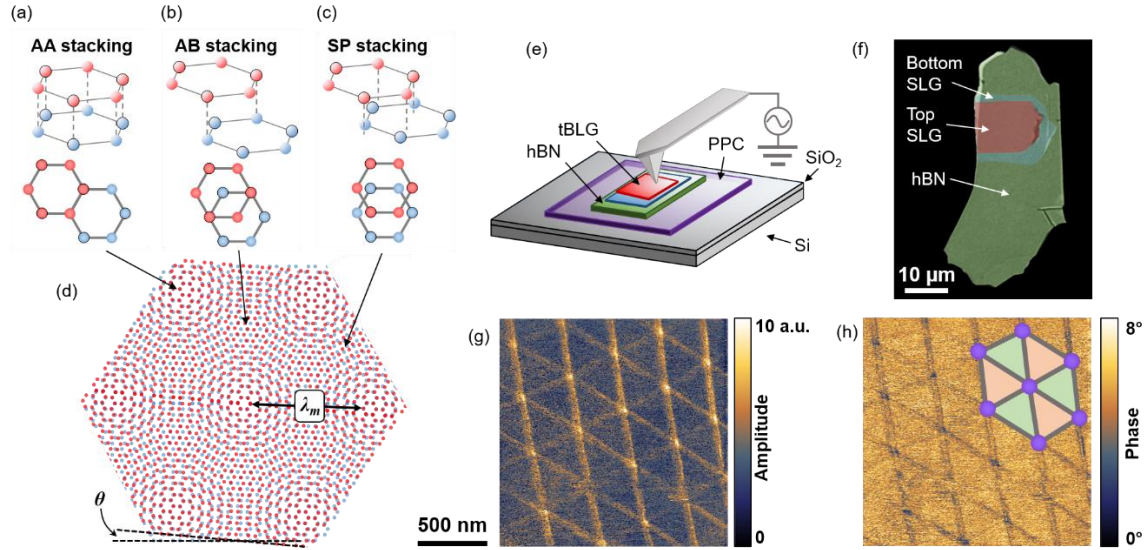

**Figure 1: Stacking order domains in twisted bilayer graphene and visualization by PFM.** The three main bilayer graphene stacking configurations are shown in (a)-(c) and their locations within a moiré pattern with a twist of angle $\theta$ and a wavelength of $\lambda_m$ (d). The PFM experiment and the sample geometry of a typical setup is shown in (e) and an optical microscopy image (f) with colour-coded areas showing the vdW heterostructure. The moiré pattern superlattice is revealed through PFM amplitude (g) and phase (h) images. The overlay shows the stacking domains; AA sites in purple, AB/BA domains in green and orange and DWs in grey.

Regardless of the origin of the contrast obtained here, the technique is beneficial for fast and simple identification of moiré domain patterns in twisted bilayer systems. PFM can provide invaluable information on the local twist angle which is otherwise hard to achieve. We applied this technique to a number of materials systems for which examples are given in Figure 2. PFM proved capable of revealing the moiré patterns in multiple systems with astounding resolution down to moiré wavelengths below 5 nm (Figure 2j,k,l) and over several micron length scales (Figure S2) limited only by sample size. This imaging technique is not limited to semi-metallic graphene but is also clearly observed in semiconductors such as $WSe_2$ (Figure 2d,e,f) and $MoSe_2$ and insulators such as BN (Figure 2g,h,i) and their heterostructures (Figure 2a,b,c). In fact, we believe that this technique will be able to accurately map the moiré pattern of any 2D materials system. The universal applicability indicates that the underlying origin of this phenomenon is equally universal and does not depend on the detailed behavior of the constituent 2D materials.

To illuminate the nature of the electromechanically induced surface deformation we first perform a type of simple "vector" PFM [21] for our prototypical system of tBLG as is outlined in Figure 3. PFM is sensitive to out-of-plane (vertical) and in-plane (lateral) surface displacements which can be independently detected by the quadrant photodiode detector as is illustrated in Figure 3a,b. For lateral displacements perpendicular to the cantilever axis, the cantilever deforms through torsion resulting in a lateral signal on the photodiode. In contrast lateral surface displacement parallel to the axis results in cantilever flexure which is measured as a vertical deflection on the photodiode detector (Figure S3) and is more difficult to distinguish from a real vertical surface displacement. However, by rotating the sample under investigation both in-plane surface displacement vectors can be accessed, and the intrinsic vertical displacement, which should be invariant upon sample rotation can be determined. In the case of bilayer graphene, sample rotation produced a varying vertical PFM contrast that was consistent with an in-plane displacement component aligned with the cantilever axis and measured through flexure (Figure S4). This is direct evidence that PFM measures a largely in-plane response and any apparent vertical signal is dominated by the in-plane flexural crosstalk.

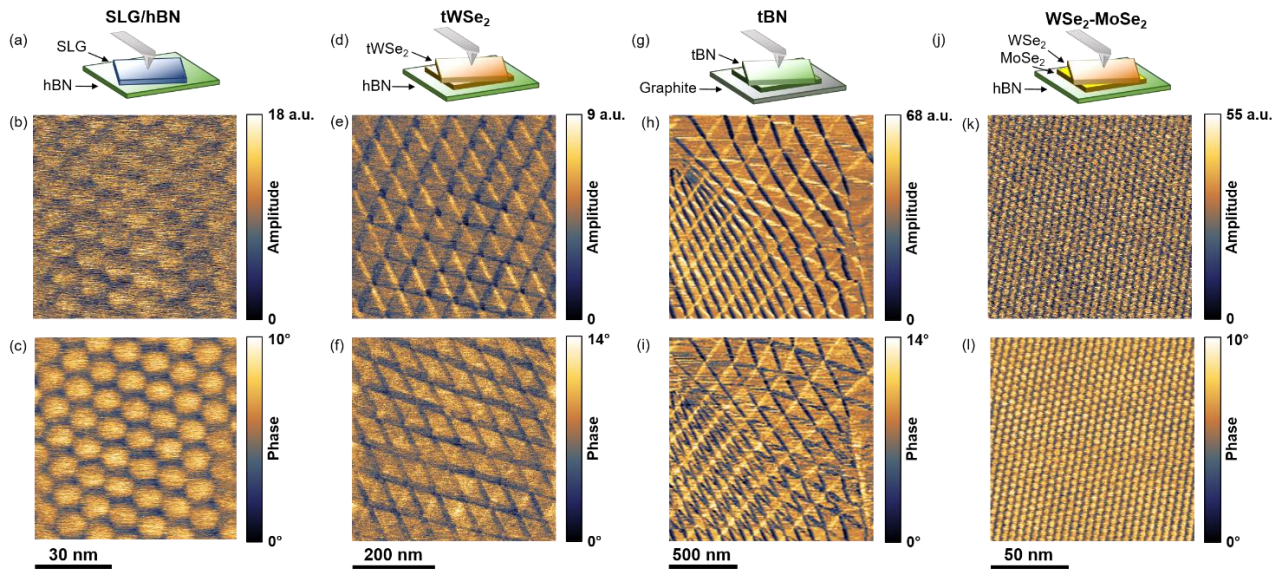

**Figure 2: Examples of PFM imaging of moiré superlattices in various vdW heterostructures.** Schematic representations (top row) corresponding PFM amplitude (middle row) and phase (bottom row) of monolayer graphene on BN (a)-(c) twisted bilayer $WSe_2$ (d)-(f) twisted bilayer BN (g)-(i) and twisted $WSe_2$-$MoSe_2$ heterobilayer (j)-(l).

Figure 3d shows the in-plane PFM amplitude measured across domain walls as a function of the domain wall angle relative to the cantilever axis. The measured lateral signal is a minimum/zero when the domain wall is aligned with the cantilever axis and reaches a maximum when the domain wall is perpendicular to the cantilever axis. This indicates that the in-plane surface deformation is parallel to the domain wall. For this reason, only 2 sets of domain walls are seen in a single scan. This is clearly seen for two images of the same moiré pattern with the sample rotated by 90°, where domain walls perpendicular to the axis show high contrast and those parallel show weak contrast. However, recombining two images from orthogonal scans should be sufficient to reconstruct the original in-plane surface displacement, as depicted schematically in Figure 3c. Figure 3g shows the reconstructed image derived from Figures 3e and 3f, which closely agrees with the expected moiré pattern. From this analysis we can surmise that the applied out-of-plane electric field leads to an in-plane surface deformation that must be aligned along the length of the domain wall.

We next investigate the origin of the apparent electromechanical displacement oriented along the domain wall, focusing on the case of tBLG: since all the studied systems exhibit similar features in PFM originating from moiré patterns, our analysis can also be extended to them. We first note that in tBLG both AA (P6/mmm, point group 6/mmm) and AB (P$\bar{3}$m1, point group $\bar{3}$m) stacking should not have a piezoelectric response. However, this symmetry is necessarily broken as the stacking changes from AB in the middle of the domains to AA and SP, which is always accompanied by formation of large strain gradients.

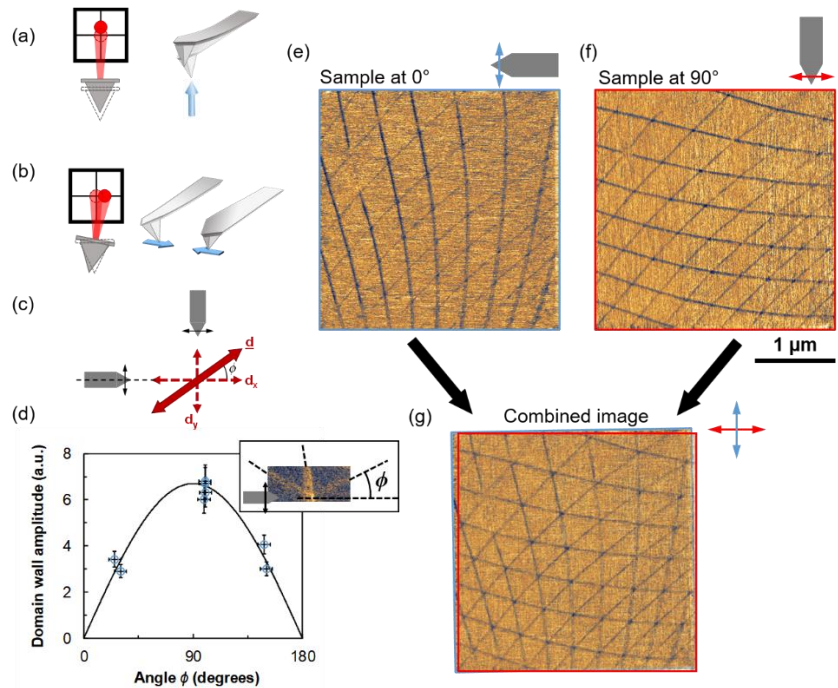

**Figure 3: PFM imaging modes, cantilever dynamics and the resulting effects on contrast in tBLG.** A vertical displacement of the cantilever due to an out-of-plane surface deformation (blue arrow) leads to a vertical deflection on the photodiode detector (a). In-plane surface deformation (blue arrows) lead to torsional bending of the cantilever and a resulting lateral deflection on the photodiode detector (b). For an in-plane surface deformation vector $d$, detection for lateral PFM will only measure the component that is oriented perpendicular to the cantilever axis denoted as $d_x$ (cantilever: grey pointed rectangle. Axis: black dotted line. Measurement sensitivity direction: black double-headed arrow) as seen in (c). In order to measure the orthogonal component, $d_y$, physical rotation of the sample is required so that reconstruction of the total in-plane surface displacement $\underline{d}$, is possible through "vector" PFM. A measure of the apparent domain wall amplitude as a function of the angle separation between the cantilever axis and the domain wall (inset) shows results (blue circles) consistent with a sine function (solid black line) in (d). (e)-(g) Simple vector PFM performed by sample rotation of ~90°; phase images at 0° (e) and 90° (f) can be recombined to reconstruct the full moiré pattern (g).

Generally, the symmetry-breaking as a result of the strain gradients can allow piezoelectric coupling to the out-of-plane field. This coupling of the piezoelectric response $e_{\alpha\beta\gamma}$ to the strain gradients $\frac{\partial \epsilon_{\lambda\mu}}{\partial x_\kappa}$ is described by a six-rank tensor $T_{\alpha\beta\gamma\kappa\lambda\mu}$:

$$e_{\alpha\beta\gamma} = T_{\alpha\beta\gamma\kappa\lambda\mu} \frac{\partial \epsilon_{\lambda\mu}}{\partial x_\kappa} \tag{1}$$

Where $e_{\alpha\beta\gamma}$ is defined as the stress $\sigma_{\beta\gamma}$ generated by applying electric field $E_\alpha$:

$$\sigma_{\beta\gamma} = E_\alpha e_{\alpha\beta\gamma} \tag{2}$$

Strain gradients and their responses to electric fields can also be linked to polarization via the flexoelectric effect [22]. Here, a strain gradient gives rise to an electric polarization:

$$P_\alpha = \mu_{\alpha\beta\gamma\lambda} \frac{\partial \epsilon_{\gamma\lambda}}{\partial x_\beta} \tag{3}$$

This polarization then gives rise to a piezoelectric effect under the influence of an electric field. An alternative to the flexoelectric effect is a direct coupling of strain gradients to piezoelectric stresses, without the necessity for the existence of a polarization in the material. We will quantify the magnitude of both of these effects for twisted bilayer graphene below, and then make general remarks about other materials.

We start with the flexoelectric effect. For $\bar{3}$m and 6/mmm point groups, there are 7 independent flexoelectric coefficients. For graphene and bilayer graphene at small displacement field, the large in-plane conductivity will screen lateral polarizations, so we only need to consider out-of-plane polarizations generated via the flexoelectric effect, i.e. the coefficients $\mu_{zz,xx}$, and $\mu_{zx,zx}$ (see Figure S5 for schematics). Figure 4a,b gives a physical view of the origin of this polarization; flat graphene sheets have planar σ-bonds and symmetric π-bonds out of the plane. However, when a curvature is present the bonds bend away from purely planar in character to possess a component of sp$^3$ bonding (Figure 4b) as opposed to the purely sp$^2$ case of flat graphene (Figure 4a). This causes an asymmetric distribution of the electron orbitals and hence gives rise to a polarization [23-26]. It is challenging to directly model the polarization at a domain wall in twisted bilayer graphene; instead, to estimate the magnitude of these coefficients, we perform calculations on carbon nanotubes (Figure 4c). As demonstrated in Ref. [23,24], the curvature of the nanotube induces a polarization, which scales linearly with the inverse radius of the nanotube, i.e., the gradient of strain. This is manifested as a potential change over the nanotube which is calculated (see Methods for computational details) along the radial line shown in the inset to Figure 4c and plotted for a range of nanotubes as a function of the inverse of the nanotube radius, R. In order to make a closer connection to bilayer graphene, we also plot potential differences for double-walled nanotubes in Figure 4c and find that additional layers are approximately additive in terms of the voltage drop. For our purposes, the slope of the lines in Figure 4c can be converted into an estimate for the shear flexoelectric coefficient $\mu_{zx,zx}$ [see Supplementary Section

S5 for details]. Using the bilayer results from Figure 4c we obtain $\mu_{zx,zx} \simeq -0.03$ nC/m. This is of the same order as the shear clamped-ion flexoelectric coefficients for example, in perovskite oxides [27,28]. Therefore, if we assume 70pm buckling of the monolayer over 10nm [10] we obtain $P_z = 0.002$ μC/cm². This result is low compared to typical values found in ferroelectric materials where polarization is usually tens of microcoulomb per centimetre squared e.g. in $BaTiO_3$, $P_z = 25$ μC/cm². However, the exact value is likely to strongly depend on the local strain gradient which may not be uniform across the domain wall or AA site.

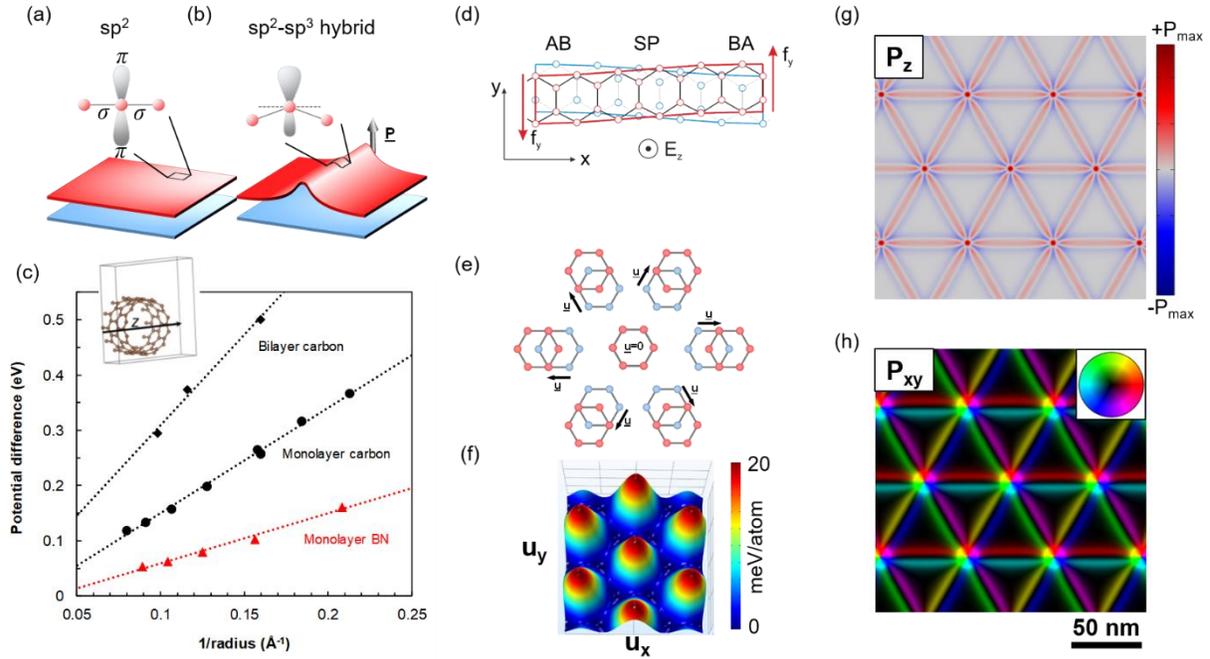

**Figure 4: Strain-gradient and curvature induced polarization.** In tBLG, a strain gradient leads to a bending of the σ-bonds from in-plane sp² (a) to a mixed in-plane/out-of-plane sp²-sp³ character (b) due to the curvature present at the domain walls; the asymmetric π-orbitals give rise to an out-of-plane polarization, **P** (black arrow). (c) Nanotubes of various materials and radii hence, strain, develop a potential difference from inside to outside of the nanotube along its radius (inset). (d) A schematic of stacking order across a domain wall and the associated piezoelectric forces $\pm f_y$ due to an out-of-plane applied electric field $E_z$. (e) The low energy lateral shifts of a top graphene layer (red) relative to a lower layer (blue) gives AB-type stacking with 6-fold symmetry and allows definition of the order parameter **u**. The Landau free energy landscape is shown in (f) with u=0 at the local maxima. Calculation of flexoelectric coupling gives rise to out-of-plane polarization $P_z$ localized at the highest strain-gradient regions, i.e. AA-sites and domain walls (g) while in-plane polarizations $P_{xy}$ emerge in the vicinity of the domain walls with polar vectors of opposite direction and vortices surrounding the AA-sites (h). Colour denotes direction of $P_{xy}$ and strength gives magnitude.

We next consider the direct coupling of strain gradients to piezoelectric stresses. In Supplementary Section S6 we list the symmetry-allowed piezoelectric responses as a result of this coupling. In particular, we consider the components:

$$e_{zyz} = C_1 \frac{\partial \epsilon_{xy}}{\partial x}, \qquad (4) \qquad e_{zxy} = C_2 \frac{\partial \epsilon_{xy}}{\partial z} \qquad (5)$$

where *x* is the direction normal to the wall. The first expression here describes the shear piezoelectric coefficient $e_{zyz}$ in relation to the shear gradient $\frac{\partial \epsilon_{xy}}{\partial x}$, closely tied to the flexoelectric effect described by coefficient $\mu_{zx,zx}$. In contrast, the second term presents an entirely new coupling, given by the difference of the shear strain at top and bottom layers; in effect, a direct strain-gradient-induced piezoelectricity which leads to the observed PFM contrast without induction of polarization. In order to estimate the magnitude of these effects, we conducted density functional theory (DFT) calculations of a simplified model of a domain wall similar to those in tBLG (i.e., where the stacking goes from AB→SP→BA→SP). We start by setting an out-of-plane corrugation of the top graphene layer, modeling the situation described in Supplementary Section S5 and apply an out-of-plane electric field to the system via placing a dipole in the vacuum region of the supercell. We then introduce in-plane shear strains and strain gradients (details in Supplementary Section S6). In the latter case, application of the out-of-plane electric field leads to appearance of forces acting on the top layer along the wall along with elastic stresses that are roughly proportional to $\frac{\partial \epsilon_{xy}}{\partial x} E_z$, (Figure 4d) agreeing with our symmetry considerations. Analyzing the results of our DFT simulations, we can estimate (see Supplementary Section S6 for details) that a measureable surface displacement of the domain wall of 0.1-1 pm is expected at applied electric fields of 10-100MV/cm, which, while high, is not unreasonable since electric fields may reach large values near the tip apex. We have therefore shown a fundamental possibility of surface displacement along the wall induced by an out-of-plane electric field in free-standing bilayer graphene (Figure 4d). In reality, other mechanisms can contribute to the piezoelectric response, such as substrate interactions or bandgap opening but which are beyond the scope of this section.

Both of the mechanisms described above in general contribute to the observed PFM contrast in the various materials studied. In the case of semiconducting or insulating systems, in-plane polarizations can also be sustained which can give rise to larger in-plane piezoelectric response. To gain further insight into the piezoelectric response and flexoelectric coupling for a generic insulating material, we conducted finite-element simulations of the polarization developed for various stacking configurations between the two layers of a bilayer. A given stacking configuration is defined by the relative lateral translation of the two layers **u** with u=0 corresponding to AA stacking (see Figure 4e for schematic stackings). The energetic landscape as a function of order parameter is displayed in Figure 4f (see Methods for details). For each of these configurations, we calculate the out-of-plane (Figure 4g) and in-plane (figure 4h) polarizations respectively. The out-of-plane polarization is maximized at the AA site with smaller responses along the domain wall, while the in-plane response shows polar vorticity at the AA sites and strong response along the domain walls.

Strain gradients are an inherent part of moiré superlattices and here we have shown that in the regions where these are to be found there are previously undiscovered physical phenomena. In particular, the existence of polarizations in moiré structures has consequences for both the electronic and optical properties of these materials. Electronically, the presence of a polarization implies a strong modification of wavefunction extent within a moiré site, the dielectric screening properties and consequently the magnitude of Coulomb interactions. Optically, the presence of dipole moments in the moiré structures will modify the optical response both qualitatively (via selection rules) and quantitatively (via the dielectric properties of the system). A complete low-energy theoretical analysis of these issues and exploration of their consequences for the optoelectronics properties remains an open problem.

# Methods

### Sample fabrication

Samples were prepared using the standard polymer stamp dry-transfer technique on a modified optical microscope with heating stage and rotation stage. A glass slide with a polydimethylsiloxane (PDMS) stamp coated in a thin polypropylene carbonate (PPC) film is brought into contact with individual flakes previously exfoliated onto $SiO_2$/Si via standard methods. In this way a thick BN flake is first picked up and then placed in contact with approximately half of a large graphene monolayer. During pick-up the graphene monolayer tears along the edge of the BN. The desired twisted angle is then made through rotational misalignment of the picked-up partial graphene monolayer and the remaining portion of the graphene on $SiO_2$/Si. After the second graphene pick-up the PPC is carefully removed from the PDMS stamp and placed onto a $SiO_2$/Si chip (heated for better adhesion). Other material samples e.g. $WSe_2$ are fabricated in a similar manner.

### Piezoresponse force microscopy

PFM was performed on a Bruker Dimension Icon with a Nanoscope V Controller. Typically, Oxford Instrument Asylum Research ASYELEC-01 Ti/Ir coated silicon probes a force constant of ~3 $Nm^{-1}$ were used. Generally, AC bias magnitudes were <1 V with resonance frequencies in the range of ~300 kHz for vertical and 750-850 kHz for lateral PFM. Single frequency excitation at the resonance peak was found to give stable imaging conditions as the peaks did not shift appreciably (i.e. <500 Hz) as surface roughness on such atomically flat surfaces is minimal. Low loading forces of less than approx. 50 nN were routinely used and produced the best imaging while also maintaining good tip apex quality. Amplitude values are displayed in arbitrary units as calibration is challenging and unreliable for comparison between different cantilevers. Results were confirmed on an Oxford Instruments Asylum Research Cypher AFM operating in the dual-AC resonance tracking mode (Figure S7).

### Density Functional Theory

Density functional theory (DFT) calculations of nanotubes are performed using the PBE generalized gradient functional [29] and projector-augmented wave [30] implemented in the VASP package [31]. A Monkhorst-Pack [32] k-mesh of 10×1×1 (with 10 points in the direction parallel to the nanotube) and an energy cutoff for the plane-wave basis set of 500 eV was used. No atoms were allowed to relax in the calculations; the C-C (B-N) distance was fixed to 1.42 (1.45) Å.

Calculations of the tBLG piezoelectric response are done in the framework of the local density approximation (LDA) to DFT with the in-house code LAUTREC. Atomic cores are represented by norm-conserving pseudopotentials in the Troullier-Martins form. The bilayer structures are built by placing two graphene sheets at a fixed distance of 6.3 bohr, close to the calculated equilibrium distance within LDA. The out-of-plane cell parameter is set to 30 a.u., which provides enough vacuum to decouple the system from its periodically repeated images. Each sheet is distorted by a sinusoidal acoustic wave, where the amplitude in the two layers has opposite signs. The amplitude is set in such a way that maxima and minima correspond to AB or BA stacking, respectively. No further relaxation is considered. The Brillouin zone of the 16-cell geometry is sampled by a 1x4x36 Monkhorst-Pack grid of k-points (for smaller/larger geometries we used equivalent or denser grids). We use a plane-wave cutoff of 80 Ry. The out-of-plane electric field is applied by introducing a suitable external dipole layer within the vacuum region.

**Finite-element simulations**

The following energy functional $W$ was used for finite-elements calculations:

$$W = V(\mathbf{u}) + W_{\text{elast}},$$

$$W_{\text{elast}} = \frac{C_{11}}{2}(\epsilon_{11}^2 + \epsilon_{22}^2) + C_{12}\epsilon_{11}\epsilon_{22} + (C_{11} - C_{12})\epsilon_{12},$$

$$\epsilon_{11} = \frac{\partial u_1}{\partial x}, \qquad \epsilon_{22} = \frac{\partial u_2}{\partial y}, \qquad \epsilon_{12} = \frac{1}{2}\left(\frac{\partial u_1}{\partial y} + \frac{\partial u_2}{\partial x}\right),$$

$$C_{11} = \frac{E^{2D}}{1-\nu^2}, \qquad C_{12} = \frac{\nu E^{2D}}{1-\nu^2},$$

where $V(\mathbf{u})$ is the 6-fold-symmetric periodic potential due to van-der-Waals interlayer interaction shown in Figure 4f [9,33], $W_{\text{elast}}$ is the elastic energy of the top layer, and $\epsilon_{ij}$ is the 2D elastic strain of the top layer. For our simulations, we use Young modulus $E^{2D} = 340$ N/m and Poisson ratio $\nu = 0.3$ [9].

In our simulations, we set the initial distribution of the order parameter $(u_1, u_2) = (\theta y, -\theta x)$, corresponding to the rotation of the top graphene layer by a small angle $\theta$, and let the system relax until the minimal energy is reached. As a result, the domain structure is reached with $a_0/\theta$, with clearly defined domain walls and vortices.

## Acknowledgements

This work is supported by the Programmable Quantum Materials (Pro-QM) programme at Columbia University, an Energy Frontier Research Center established by the Department of Energy (grant DE-SC0019443). L.J.M. acknowledges support from the Swiss National Science Foundation (grant no. P400P2_186744). Synthesis of MoSe$_2$ and WSe$_2$ was supported by the National Science Foundation Materials Research Science and Engineering Centers programme through Columbia in the Center for Precision Assembly of Superstratic and Superatomic Solids (DMR-1420634). The Flatiron Institute is a division of the Simons Foundation. M.S. and K.S. acknowledge the support of the European Research Council under the European Union's Horizon 2020 research and innovation program (Grant Agreement No. 724529), Ministerio de Economia, Industria y Competitividad through Grant Nos. MAT2016-77100-C2-2-P and SEV-2015-0496, and the Generalitat de Catalunya (Grant No. 2017SGR 1506). We thank Drew Griffin and Tim Walsh from Oxford Instruments Asylum Research for confirmation of PFM results.

# Supplementary Information

**S1. Corresponding topography for piezoresponse force microscopy (PFM)**

Figure S1 shows the height images associated with the PFM amplitude and phase images presented in Figures 1, 2 of the main text. No clear topography features correlate with the respective moiré patterns.

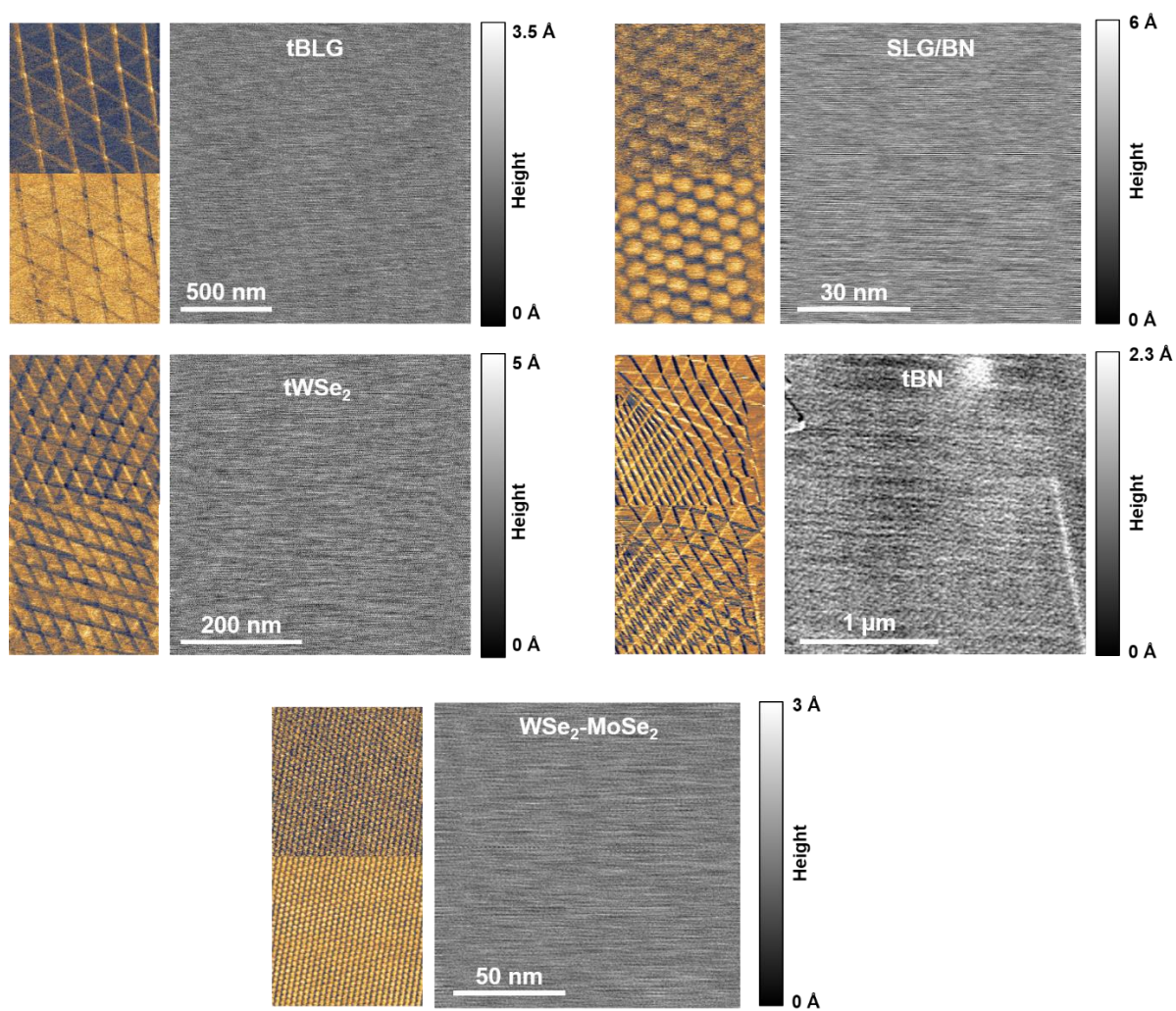

**Figure S5 Topography of various materials investigated by PFM.** The PFM amplitude and phase images (reduced size) are found to the left of the corresponding topography images with colour scales.

## S2. Large-scale PFM imaging of moiré superlattices

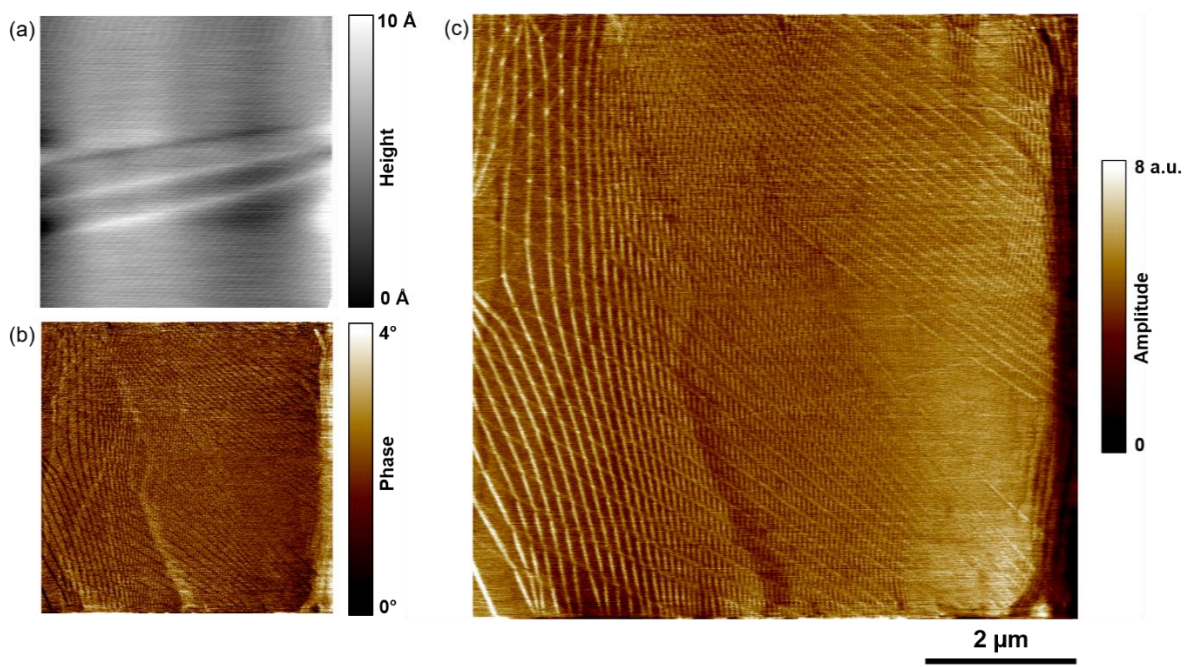

**Figure S2 Example of large-scale mapping of moiré superlattice.** A PFM image for a scan of 8×8 µm$^2$; topography (a), phase (b) and amplitude (c) demonstrating that this technique can be used to observe the moiré across length scales that span orders of magnitude. Note the huge variation in moiré wavelength from ~500 to ~50 nm and presence of large strains.

## S3. Cantilever flexure

In Figure S3 we show the two simple modes of cantilever oscillation that can lead to a vertical deflection on a quadrant photodiode detector. Cantilever flexure from in-plane surface displacement components oriented along the cantilever axis gives a vertical deflection that can be erroneously measured as a genuine out-of-plane surface displacement.

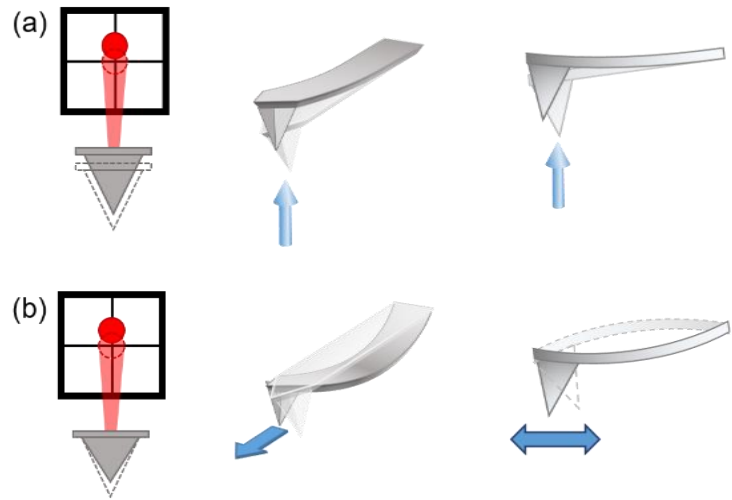

**Figure S3 Cantilever flexural mode.** A simple schematic explanation of how out-of-plane (a) and in-plane (b) surface displacements give rise to an indistinguishable vertical deflection.

## S4. Vector PFM on exfoliated graphene

Figure S4 shows a flake of as-exfoliated few-layer graphene on $SiO_2$ with PFM performed at three angles of sample rotation 0°, 90° and 180°. First of all, it is clear that the vertical images change with rotation clearly indicating that the response is not due to a purely out-of-plane surface displacement. Secondly if we cross-examine the vertical images with the lateral images with a relative sample rotation of 90° we see very similar or close to identical images. This tells us that the vertical images are dominated by the flexural mode of the cantilever dynamics as outlined in Figure S3 and thus is sensitive to the in-plane surface displacements.

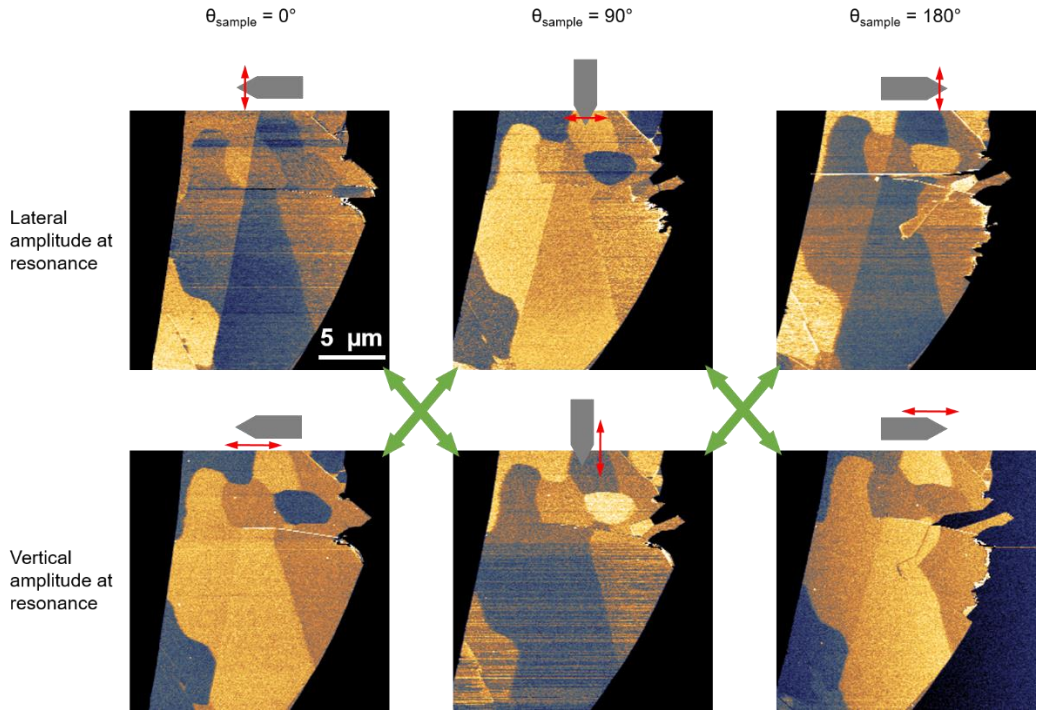

**Figure S4 Vector PFM of as-exfoliated few-layer graphene.** PFM amplitude images obtained at third sample rotations and at both vertical and lateral resonances sequentially. The gray pointed rectangle denotes the orientation of the cantilever with respect to the flake and the red double headed arrow gives the effective measurement sensitivity direction. The green double-headed arrows show images with near-identical contrast and hence the same direction of sensitivity as given by the red double-headed arrows.

## S5. Flexoelectricity

Flexoelectricity is the appearance of a polarization in response to a strain gradient. In the main text we consider a reduced set of strain gradients relevant to our experimental setup. Figure S5 shows schematic depictions of strain gradients that can give rise to polarizations and their respective flexoelectric coefficients.

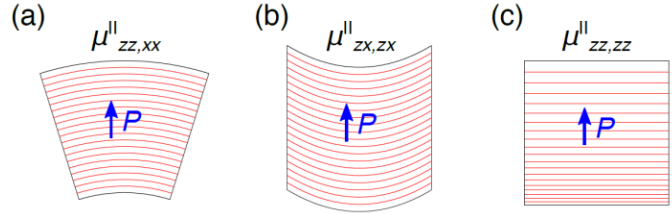

**Figure S5 Flexoelectric coefficients and the relevant strain gradients that give rise to them.** For the case of twisted bilayer vdWs heterostructures the out-of-plane strain gradient depicted in (c) is not relevant as this would require at least 3 layers.

In order to estimate the polarization of nanotubes of various radii as shown in Figure 4c of the main text we use the "flexocoupling coefficient," [28] which can be written as

$$\Phi_{zx,zx} = -\lim_{t\to\infty} \frac{1}{t}\frac{\partial \Delta V}{\partial \epsilon_{zx,x}} = \frac{1}{\varepsilon_{zz}}\mu_{zx,zx} \tag{S5}$$

where $t$ is the thickness of the film and $\varepsilon_{zz}$ corresponds to, in our case, the clamped-ion dielectric constant perpendicular to the film thickness. We can clearly see the correspondence between bending of the layers and the shear coefficient $\mu_{zx,zx}$, and $1/R$ gives the gradient $\epsilon_{zx,x}$. Therefore, the slope of the lines correspond to $\frac{\partial \Delta V}{\partial \epsilon_{zx,x}}$. Using the bilayer results from Figure 4a (which are for $t = 4:70$ Å, and assuming a dielectric constant of $\varepsilon_{zz} = 6\varepsilon_0$ [2]), we obtain $\mu_{zx,zx} \simeq -0.03$ nC/m which is the value quoted in the main text and used as a basis for the polarization estimate.

## S6. Piezoelectric response by strain gradients: phenomenological theory and first-principles simulations

The stress-piezoelectric tensor $e_{\alpha\beta\gamma}$ describing the stress $\sigma_{\beta\gamma}$ generated by applying an electric field $E_\alpha$ is zero in bilayer graphene with AB-stacking due to the presence of inversion symmetry generator in the $\bar{3}m$ point group of the system. The piezoelectric effect may become non-zero if coupled to other order parameters in the system that inevitably change their values in domain walls and in vortices. As discussed in the main text, coupling to the polarization in bilayer graphene with in-plane conductivity can only lead to non-zero $e_{zxx}$, $e_{zyy}$ and $e_{zzz}$ components, which cannot give a PFM contrast along the wall. The lowest-order coupling to strain would be described by a fifth-rank tensor which is, again, zero due to the symmetry restrictions of the point group of the bilayer graphene.

On the other hand, the coupling of $e_{\alpha\beta\gamma}$ to the strain gradients $\frac{\partial \epsilon_{\lambda\mu}}{\partial x_\kappa}$ is perfectly allowed, and is described by a six-rank tensor $T_{\alpha\beta\gamma\kappa\lambda\mu}$:

$$e_{\alpha\beta\gamma} = T_{\alpha\beta\gamma\kappa\lambda\mu} \frac{\partial \epsilon_{\lambda\mu}}{\partial x_\kappa}.$$

Focusing only on the piezoelectric responses to the out-of-plane electric field $E_z$, we can write down all the components allowed by $\bar{3}m$ symmetry (the result is the same for 6/mmm symmetry of AA stacking):

$$e_{zxx} = a_1 \left( \frac{\partial \epsilon_{xz}}{\partial x} + \frac{\partial \epsilon_{yz}}{\partial y} \right) + a_2 \frac{\partial (\epsilon_{xx} + \epsilon_{yy})}{\partial z} + a_3 \frac{\partial \epsilon_{zz}}{\partial z} + b_1 \left( \frac{\partial \epsilon_{xz}}{\partial x} - \frac{\partial \epsilon_{yz}}{\partial y} \right) + b_2 \frac{\partial (\epsilon_{xx} - \epsilon_{yy})}{\partial z},$$

$$e_{zyy} = a_1 \left( \frac{\partial \epsilon_{xz}}{\partial x} + \frac{\partial \epsilon_{yz}}{\partial y} \right) + a_2 \frac{\partial (\epsilon_{xx} + \epsilon_{yy})}{\partial z} + a_3 \frac{\partial \epsilon_{zz}}{\partial z} - b_1 \left( \frac{\partial \epsilon_{xz}}{\partial x} - \frac{\partial \epsilon_{yz}}{\partial y} \right) - b_2 \frac{\partial (\epsilon_{xx} - \epsilon_{yy})}{\partial z},$$

$$e_{zxy} = b_1 \left( \frac{\partial \epsilon_{yz}}{\partial x} + \frac{\partial \epsilon_{xz}}{\partial y} \right) + 2b_2 \frac{\partial \epsilon_{xy}}{\partial z},$$

$$e_{zzz} = a_4 \left( \frac{\partial \epsilon_{xz}}{\partial x} + \frac{\partial \epsilon_{yz}}{\partial y} \right) + a_5 \frac{\partial (\epsilon_{xx} + \epsilon_{yy})}{\partial z} + a_6 \frac{\partial \epsilon_{zz}}{\partial z},$$

$$e_{zxz} = a_7 \frac{\partial (\epsilon_{xx} + \epsilon_{yy})}{\partial x} + a_8 \left( \frac{\partial (\epsilon_{xx} - \epsilon_{yy})}{\partial x} + 2 \frac{\partial \epsilon_{xy}}{\partial y} \right) + a_9 \frac{\partial \epsilon_{zz}}{\partial x} + a_{10} \frac{\partial \epsilon_{xz}}{\partial z},$$

$$e_{zyz} = a_7 \frac{\partial (\epsilon_{xx} + \epsilon_{yy})}{\partial y} + a_8 \left( -\frac{\partial (\epsilon_{xx} - \epsilon_{yy})}{\partial y} + 2 \frac{\partial \epsilon_{xy}}{\partial x} \right) + a_9 \frac{\partial \epsilon_{zz}}{\partial y} + a_{10} \frac{\partial \epsilon_{yz}}{\partial z},$$

Here, terms with $a_i$ correspond to the effects analogous to flexoelectricity, and $b_i$ corresponds to the new effect: neither of $\left(\frac{\partial \epsilon_{xz}}{\partial x} - \frac{\partial \epsilon_{yz}}{\partial y}\right)$, $\frac{\partial(\epsilon_{xx}-\epsilon_{yy})}{\partial z}$, $\left(\frac{\partial \epsilon_{yz}}{\partial x} + \frac{\partial \epsilon_{xz}}{\partial y}\right)$, $\frac{\partial \epsilon_{xy}}{\partial z}$ can produce a non-zero flexoelectric polarization even in insulating systems, but nevertheless each of them can give a PFM response.

It is instructive to analyze which piezoelectric tensor components are active at different types of domain walls. Let us choose the coordinate system in which $x$-direction is normal to the wall, and $y$ is along the wall. Taking into account that in this case $\frac{\partial \epsilon_{yz}}{\partial x} = \frac{\partial \epsilon_{xy}}{\partial z}$, and neglecting responses from $\frac{\partial \epsilon_{zz}}{\partial z}$, $\frac{\partial \epsilon_{yz}}{\partial z}$ (because at least three layers are needed for such gradients to develop), the piezoresponses can be rewritten in a simpler form:

$$e_{zyz} = C_1 \frac{\partial \epsilon_{xy}}{\partial x}, \qquad e_{zxy} = C_2 \frac{\partial \epsilon_{xy}}{\partial z}, \qquad e_{zxz} = C_3 \frac{\partial \epsilon_{zz}}{\partial x} + C_4 \frac{\partial \epsilon_{xz}}{\partial z} + C_5 \frac{\partial \epsilon_{xx}}{\partial x},$$

$$e_{zxx} = C_6 \frac{\partial \epsilon_{xz}}{\partial x} + C_7 \frac{\partial \epsilon_{xx}}{\partial z}, \qquad e_{zyy} = C_8 \frac{\partial \epsilon_{xz}}{\partial x} + C_9 \frac{\partial \epsilon_{xx}}{\partial z}, \qquad e_{zzz} = C_{10} \frac{\partial \epsilon_{xz}}{\partial x} + C_{11} \frac{\partial \epsilon_{xx}}{\partial z}.$$

Thus, only shear domain walls, developing in-plane shear strains $\epsilon_{xy}$, can result in piezoelectric responses along the wall, given by constants $C_1$ and $C_2$.

To estimate coefficients $C_1$ and $C_2$, we perform a series of first-principles simulations using ABINIT package. We fix the displacements of the carbon atoms of the two hexagonal sublattices as follows:

$$u_y^{\text{bottom}}(x) = \pm u_y^{\text{top}}(x) = \frac{a_0}{2} \sin\left(\frac{2\pi x}{L}\right), \qquad L = N \times \sqrt{3} a_0, \qquad N = 12, 16, 32. \qquad (S6)$$

Here, minus sign corresponds to the AB→SP→BA→SP→AB stacking (see Fig. S6a). The piezoelectric response to an applied $z$-directed electric field in this case is $e_{zxy} = C_2 \frac{\partial \epsilon_{xy}}{\partial z} \approx -\frac{2C_2}{h_0} \epsilon_{xy}^{\text{bottom}}$, where $h_0$ is the interlayer distance, and the net force acting on a pair of the carbon atoms of one layer is $F_y^{\text{bottom}} = F_y^{\text{top}} = \frac{\sqrt{3}\pi^2 C_2 a_0 E_z}{N^2} \sin\left(\frac{2\pi x}{\sqrt{3} N a_0}\right)$. Therefore, $C_2$ can be extracted from the results of simulations as $C_2 = \frac{F_y^{\max} N^2}{E_z} \frac{1}{\sqrt{3}\pi^2 a_0}$, where $F_y^{\max}$ is the maximal y-component of the force acting on a pair of carbon atoms in the simulations. The plus sign in Eq. (S6) describes the non-zero net displacement of the two sublattices $u_y^{\text{top}} + u_y^{\text{bottom}} \neq 0$ (see Fig. S6c) due to, e.g., substrate interacting with the bottom layer, activating $e_{zyz} = C_1 \frac{\partial \epsilon_{xy}}{\partial x} \neq 0$ piezoelectric mechanism. The net force acting on a pair of carbon atoms of the bottom layer in this case is $F_y^{bottom} = -F_y^{top} = \frac{\sqrt{3}\pi^2 C_1 a_0 E_z}{2N^2} \sin\left(\frac{2\pi x}{\sqrt{3}N a_0}\right)$, and $C_1$ can be extracted from the simulations as $C_1 = \frac{F_y^{max} N^2}{E_z} \frac{2}{\sqrt{3}\pi^2 a_0}$.

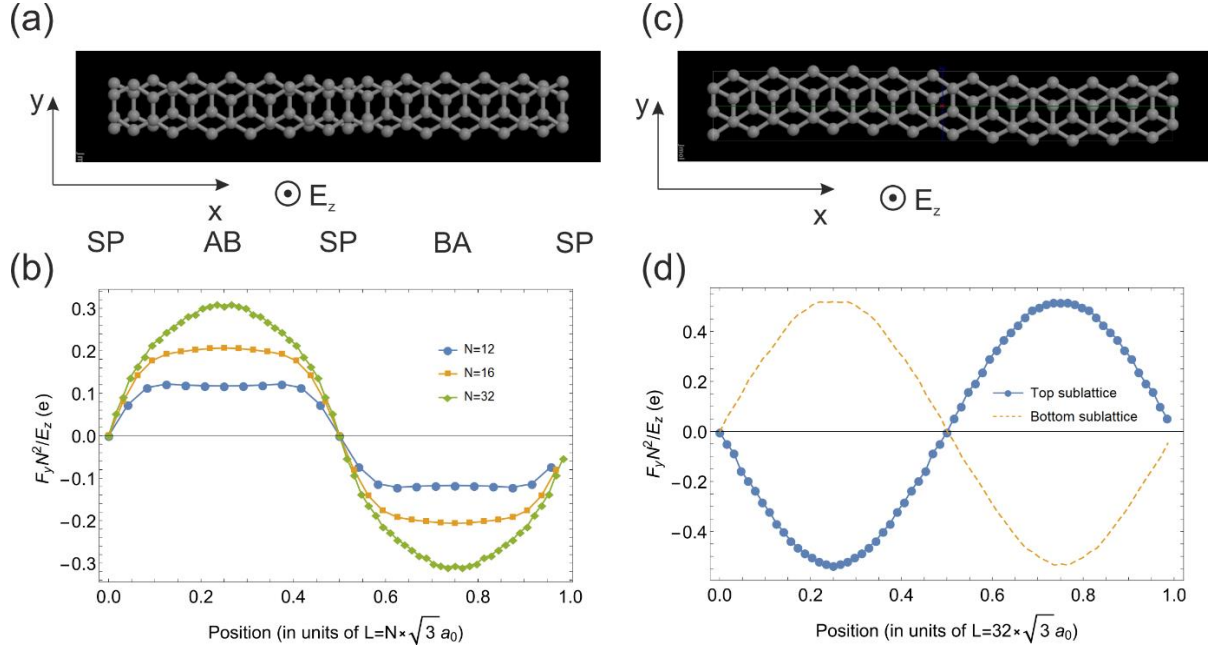

**Figure S6. Geometry of the simulated system (a,c) and the computed renormalized y-forces acting on the pair of carbon atoms in one sublattice (b,d).**

We plot in Fig. S6b,d the distributions of renormalized y-component of the force acting on a pair of carbon atoms of one sublattice $\frac{F_y N^2}{E_z}$ obtained in DFT simulations, corresponding to the geometry shown in Fig. S6a,c. As expected from the symmetry considerations, in the simulations with $u_y^{bottom} = -u_y^{top}$ (Fig. S6a) the forces acting on both sublattices are the same (Fig S6b), and in the simulations with $u_y^{bottom} = u_y^{top}$ (Fig. S6c) the forces have opposite signs (Fig. S6d). Extracting the amplitudes of modulations of the forces gives us $C_1 = 7 \times 10^{-11} \frac{C}{m}$, $C_2 = 3 \times 10^{-11} \frac{C}{m}$. For instance, applying a z-directed electric field 10 MV/cm at a domain wall of thickness $t_w \approx 10$nm would produce stresses $\sigma_{xy} \approx C_2 \frac{2\epsilon_{xy}^{top}}{h_0} E_z \approx \frac{C_2 a_0 E_z}{t_w h_0} \approx 1$ MPa. In comparison, the intrinsic stress of a shear soliton is $\approx 1$ GPa. Thus, a rough estimate of the order of magnitude of the atomic displacement of the top layer at application of the electric field 10 MV/cm is $\frac{1 \text{ MPa}}{1 \text{ GPa}} a_0 \approx 0.1$pm.

## S7. Dual AC frequency resonance tracking (DART) PFM

In order to verify that the observed PFM contrast was not instrument dependent we repeated our results of imaging the moiré superlattice of monolayer graphene on boron nitride using an Asylum Research Cypher AFM. This microscope uses a technique called DART to track the resonances of the vertical and lateral modes. From analysis of the resonance image we see that the variation in contact resonance is ~320 kHz and only varies by ~500 Hz or a value of 0.16%.

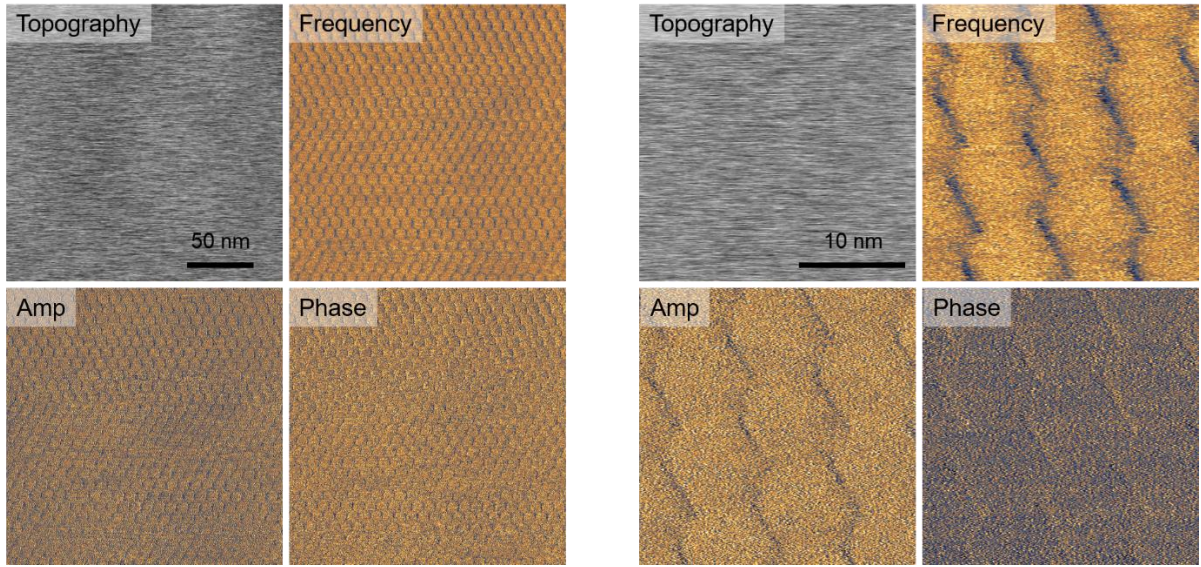

**Figure S7 DART-PFM of SLG/BN moiré superlattices.** Verification of observed PFM contrast using the DART technique to track contact resonance peaks. The right set of images shows a zoom-in from the region of the left-hand images.